\RequirePackage[2020-02-02]{latexrelease}
\documentclass[aps, prb, superscriptaddress, twocolumn, longbibliography, superscriptaddress, nofootinbib]{revtex4-2}
\usepackage[pdftex]{graphicx, color}
\usepackage{amsmath, amssymb, amscd, latexsym, bm, braket, comment}
\usepackage[colorlinks=true,pdfstartview=FitV,linkcolor=blue,citecolor=magenta,urlcolor=blue,bookmarks=true,bookmarksnumbered=true]{hyperref}

\begin{document}

\title{Analytical WKB theory for high-harmonic generation and its application to massive Dirac electrons}

\author{Hidetoshi Taya}
\email{hidetoshi.taya@riken.jp}
\address{RIKEN iTHEMS, RIKEN, Wako 351-0198, Japan}
\address{Research and Education Center for Natural Sciences, Keio University 4-1-1 Hiyoshi, Kohoku-ku, Yokohama, Kanagawa 223-8521, Japan}

\author{Masaru Hongo}
\email{masaru.hongo@riken.jp}
\address{Department of Physics, University of Illinois, Chicago, Illinois 60607, USA}
\address{RIKEN iTHEMS, RIKEN, Wako 351-0198, Japan}

\author{Tatsuhiko N. Ikeda}
\email{tikeda@issp.u-tokyo.ac.jp}
\address{Institute for Solid State Physics, University of Tokyo, Kashiwa, Chiba 277-8581, Japan}

\date{\today}

\begin{abstract}
We propose an analytical approach to high-harmonic generation (HHG) for nonperturbative low-frequency and high-intensity fields based on the (Jeffreys-)Wentzel-Kramers-Brillouin (WKB) approximation.  By properly taking into account Stokes phenomena of WKB solutions, we obtain wavefunctions that systematically include the repetitive dynamics of production and acceleration of electron-hole pairs and quantum interference due to phase accumulation between different pair production times (St\"{u}ckelberg phase).  Using the obtained wavefunctions without relying on any phenomenological assumptions, we explicitly compute electric current (including intra- and inter-band contributions) as the source of HHG for a massive Dirac system in (1+1)-dimensions under an ac electric field.  We demonstrate that the WKB approximation agrees well with numerical results obtained by solving the time-dependent Schr\"{o}dinger equation and point out that the quantum interference is important in HHG.  We also predict in the deep nonperturbative regime that (1) harmonic intensities oscillate with respect to electric-field amplitude $E_0$ and frequency $\Omega$, with a period determined by the St\"{u}ckelberg phase; (2) the cutoff order of HHG is determined by $2eE_0/\hbar \Omega^2$, with $e$ being the electron charge; and that (3) non-integer harmonics, controlled by the St\"{u}ckelberg phase, appear as a transient effect.  Our WKB theory is particularly suited for a parameter regime, where the Keldysh parameter $\gamma=(\Delta/2)\Omega/eE_0$, with $\Delta$ being the gap size, is small.  This parameter regime corresponds to intense lasers in the terahertz regime for realistic massive Dirac materials.  Our analysis implies that the so-called HHG plateau can be observed at the terahertz frequency within the current technology.
\end{abstract}

\maketitle 

{\it Introduction.}\ High-harmonic generation (HHG) is one of the most intriguing phenomena in nonlinear optics~\cite{Brabec2000}.  Owing to developments in laser technologies over the decades, HHG has been observed and analyzed in various media (e.g., atomic gases~\cite{Ferray_1988,PhysRevLett.71.1994,PhysRevA.49.2117}, liquids~\cite{Heissler2014,Luu2018}, semiconductors~\cite{ghimire2011observation, schubert2014sub,Hohenleutner2015,Luu2015,Kaneshima2018}, graphene~\cite{Mikhailov2008,Al-Naib2014,yoshikawa2017high,Hafez2018}, superconductors~\cite{matsunaga2014light,Kawakami2018,Yonemitsu2018,Nakamura2020}, strongly-correlated electrons~\cite{Murakami2018b,Imai2019,Lysne2020,Roy2020a}, and amorphous solids~\cite{you2017high, jurgens2020origin}), providing a unique opportunity to explore the fully nonperturbative regime of matter-field interaction and rich applications such as attosecond light sources and ultrafast imaging methods~\cite{PhysRevLett.115.193603}.  HHG has also been predicted in the fundamental theory of quantum electrodynamics (QED), i.e., the QED vacuum emits high harmonics when exposed to strong fields exceeding the Schwinger limit~\cite{PhysRevD.72.085005, FEDOTOV20071}.

One of the greatest theoretical challenges to elucidate the HHG mechanism is to understand nonperturbative electron dynamics under low-frequency and high-intensity fields~\cite{huttner2017ultrahigh,Ghimire2019}.  Typically, theorists numerically solve the time-dependent Schr\"{o}dinger equation (TDSE)~\cite{Wu2015,Ikemachi2017,Osika2017} (or von Neumann equation~\cite{Golde2008,Golde2011,Vampa2014,Vampa2015}) to simulate electric current and/or polarization as the source of HHG.  These studies have provided numerical evidence that the interplay between the intra- and inter-band electron dynamics plays an essential role.  Meanwhile, analytical methods are demanded to deepen the fundamental understanding and to analyze numerically inaccessible parameter regimes such as the low-frequency limit.  Analytical theories exist for the simplest single-band model~\cite{Pronin1994,ghimire2011observation}, which, however, completely neglects the interband dynamics.  For multiband models, there are Floquet-theoretical approaches~\cite{Ikeda2018b, Chinzei2020}, but analytical results have been limited to the high-frequency regime where the laser frequency $\Omega$ exceeds the band gap $\Delta$.   Thus, an analytical HHG theory has not been established for the quintessential low-frequency and high-intensity fields.

\begin{figure}[t]
\begin{flushleft}
\mbox{(a)}\\
\vspace*{15mm}
\mbox{(b)}
\vspace*{-30mm}
\end{flushleft}
\begin{center}
\includegraphics[clip, width=0.42\textwidth]{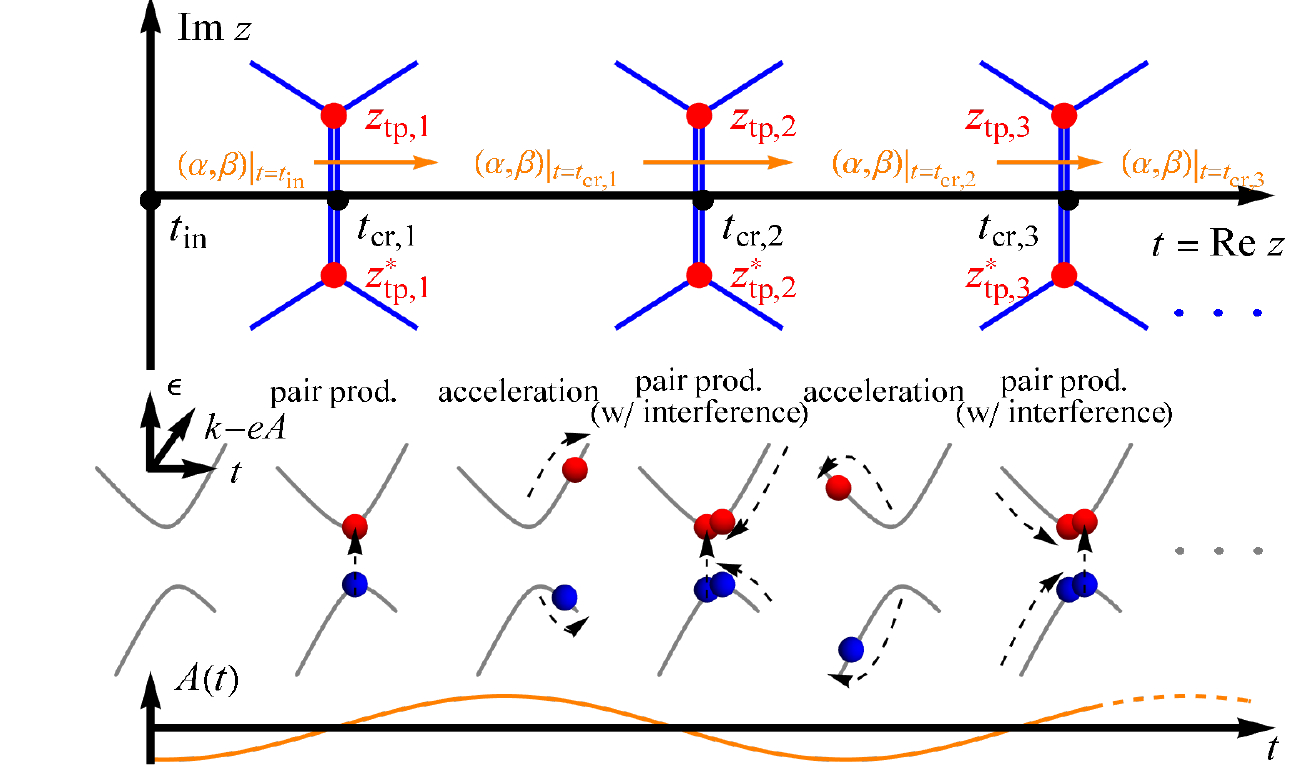}
\vspace*{-6mm}
\end{center}
\caption{\label{fig-2} (a) A typical Stokes graph, composed of Stokes lines (blue lines) and turning points (red points), and (b) the corresponding physical processes during the real-time evolution.  
}\vspace*{-5mm}
\end{figure}

In this Letter, we present an analytical method to study HHG in the nonperturbative low-frequency and high-intensity regime.  Our method is based on the (Jeffreys-)Wentzel-Kramers-Brillouin [(J)WKB~\cite{Wentzel:1926aor, Kramers:1926njj, Brillouin:1926blg, Jeffreys1924}] approximation, which has been extended by mathematicians since the 1980 (leading to the exact WKB analysis)~\cite{AIHPA_1983__39_3_211_0, CNP, AIF_1993__43_1_163_0, doi:10.1063/1.532206, AIHPA_1999__71_1_1_0, AKT1, AKT2, Aoki:1993ra} and applied successfully to various physical problems such as tunneling pair production~\cite{Enomoto:2020xlf, Enomoto:2021hfv, Taya:2020dco, Hashiba:2021npn, Sou:2021juh} and nonlinear radiative processes under strong fields~\cite{PhysRevA.99.052513, DiPiazza:2013vra, DiPiazza:2021rum}.  The WKB approximation is a semi-classical method valid in the formal limit $\hbar\to0$ or equivalently when the Keldysh parameter~\cite{Keldysh:1965ojf, Brezin:1970xf, Popov:1971ff, Taya:2014taa, Taya:2020dco}
\begin{align}
	\gamma \equiv \frac{\Omega(\Delta/2)}{eE_0} \label{Keldysh}
\end{align}
is sufficiently small ($e$ and $E_0$ are the electron charge and electric-field amplitude, respectively, and we assume $eE_0>0$ and set the speed of light to be unity).  Hence, it is suited for low-frequency and high-intensity fields, where nonperturbative processes, such as quantum tunneling  (Landau-Zener transition~\cite{landau1937theorie, zener1932non, stuckelberg1933theorie, 1932NCim....9...43M} or Sauter-Schwinger effect~\cite{Sauter:1931zz, Heisenberg:1935qt, Schwinger:1951nm}), take place.  Another advantage is that one can explicitly solve TDSE (up to some order of $\hbar$) and can compute observables directly from the wavefunction without any ad-hoc assumptions.  The wavefunction has microscopic information on the nonperturbative electron-hole dynamics under strong fields, which enables us to understand the HHG mechanism.

{\it Setup.}\ For concreteness, we consider massive Dirac electrons in (1+1)-dimensions
under a time-dependent electric field $E(t)$. The Hamiltonian reads
\begin{align}
	H \equiv \begin{pmatrix} \Delta/2 & k-eA \\ k-eA & -\Delta/2 \end{pmatrix}, \label{eq1}
\end{align}
where $A(t)\equiv-\int^t{\rm{d}}t'\,E(t')$ is the vector potential, $k$ is canonical momentum (or Bloch momentum in solids), and $\Delta$ is the gap energy.  Note that the Hamiltonian (\ref{eq1}) is relativistic and linear in $k$, so that the diamagnetic term $A^2$ is absent unlike non-relativistic Hamiltonians.  The following argument also applies to general two-level systems, e.g., Dirac materials (including QED) in other dimensions \cite{foot1}, topological insulators~\cite{Jurss2019}, semiconductors~\cite{huttner2017ultrahigh}, and graphene \cite{PhysRevLett.96.086805,McCann_2013}.  We assume that the electric field is turned on at $t=t_{\rm{in}}$, before which the system is in the ground state, and compute the induced electric current as the source of HHG.  As we will clarify soon, our WKB theory is valid in the non-perturbative regime $\gamma \lesssim 1$.  For realistic massive Dirac materials $\Delta={\mathcal O}(10\,{\rm meV})$~\cite{dirac1,dirac2}, this corresponds to intense ($E_0\gtrsim1$\,kV/cm) terahertz [$\Omega/(2\pi)\sim10^{12}$\,Hz] lasers, which are available within the current laser technology used in the terahertz HHG observations~\cite{Hafez2018,Cheng2020,Kovalev2020}.

{\it WKB solution and Stokes phenomenon.}\ To solve TDSE with the Hamiltonian (\ref{eq1}), we expand the solution $\psi$ by WKB wavefunctions $\psi_\pm$, 
\begin{align}
	\psi(t) = \alpha \psi_-(t) + \beta \psi_+(t), 
	\label{eq3}
\end{align}
where $\alpha$ and $\beta$ are called Stokes constants and satisfy $1=|\alpha|^2+|\beta|^2$ so that $1=\psi^\dagger\psi$.  We work within the lowest-order WKB approximation, where $\psi_\pm$ is given by the instantaneous eigenstates for the Hamiltonian (\ref{eq1}) with dynamical phase factors:
\begin{align}
	\psi_+
		&=
			\frac{1}{\sqrt{2}}\sqrt{1-\frac{\Delta/2}{\epsilon}}
			\begin{pmatrix} 
				+ \frac{k-eA}{\epsilon-\Delta/2} \\ 1
			\end{pmatrix} {\rm{e}}^{-\frac{\rm{i}}{\hbar} \int^t_{0} {\rm{d}}t\,\epsilon } , \nonumber\\
	\psi_-
		&=
			\frac{1}{\sqrt{2}}\sqrt{1-\frac{\Delta/2}{\epsilon}}
			\begin{pmatrix} 
				1 \\ - \frac{k-eA}{\epsilon-\Delta/2}
			\end{pmatrix} {\rm{e}}^{+\frac{\rm{i}}{\hbar} \int^t_{0} {\rm{d}}t\,\epsilon },  \label{eq-4}
\end{align}
with $\epsilon\equiv\sqrt{(\Delta/2)^2+(k-eA)^2}$ being the instantaneous eigenenergy, which includes the intraband acceleration by the external field.

To determine the Stokes constants $\alpha$ and $\beta$, we analytically continue the instantaneous energy onto the complex $z$-plane $\epsilon(t\in{\mathbb R})\to\epsilon(z\in{\mathbb{C}})$ and analyze the associated Stokes graph (see Fig.~\ref{fig-2}).  The Stokes constants $\alpha$ and $\beta$ take constant values within each Stokes region, which is defined as a region in the complex $z$-plane separated by Stokes lines ${\mathcal C}_{z^{\rm{tp}}}\equiv\{z\in{\mathbb{C}}\mid0={\rm{Im}}[\frac{\rm{i}}{\hbar} \int^z_{z^{\rm{tp}}}{\rm{d}}z\,\epsilon]\}$, with $z^{\rm{tp}}\in{\mathbb{C}}$ being turning points such that $0 \equiv \epsilon(z^{\rm{tp}})$.  For the instantaneous energy $\epsilon$ that are real on the real axis, turning points and Stokes lines are symmetric in the upper and lower complex $z$-planes, and the pairs of turning points $(z^{\rm{tp}},z^{{\rm{tp}}*})$ are connected with a doubly degenerate Stokes line crossing the real axis only once at $ t^{\rm{cr}}=\{t\in{\mathbb{R}}\,|\,0={\rm{Im}}[\frac{\rm{i}}{\hbar}\int^{t}_{z^{\rm{tp}}}{\rm{d}}z\,\epsilon]\}$~\cite{Taya:2020dco}.  Whenever $t\in{\mathbb R}$ crosses a degenerate Stokes line at $t=t^{\rm{cr}}$ during the time-evolution, the Stokes constants $\alpha$ and $\beta$ jump discontinuously (Stokes phenomenon) as~\cite{Taya:2020dco, Dumlu:2011rr}
\begin{align}
	\!\!\left.\begin{pmatrix} \alpha \\ \beta \end{pmatrix}\right|_{t^{\rm{cr}}_{n}+0^+} 
		\!\!\!\!\!\!=\! \begin{pmatrix} 1 &  \!\!\!\!(-1)^{n} {\rm{e}}^{-\frac{\sigma_n^* }{\hbar} } \\ \!-(-1)^{n} {\rm{e}}^{-\frac{\sigma_n}{\hbar} } & \!\!\!\!1 \end{pmatrix}
		\!\!\left.\begin{pmatrix} \alpha \\ \beta \end{pmatrix}\right|_{t^{\rm{cr}}_{n}-0^+} \!,\!\! \label{eq9} 
\end{align}
where $\sigma_n\equiv2{\rm{i}}\int^{z^{\rm{tp}}_n}_{0}{\rm{d}}z\,\epsilon$ (${\rm{Re}}\,\sigma_n>0$), $t^{\rm{cr}}_{n}$ ($t^{\rm{in}}<t^{\rm{cr}}_1<t^{\rm{cr}}_2<\cdots$) is the $n$-th crossing associated with $(z^{\rm{tp}}_n,z^{{\rm{tp}}*}_n)$, and we neglected subleading ${\mathcal{O}}({\rm{e}}^{-\frac{2}{\hbar}{\rm{Re}}\,\sigma_n})$ terms by assuming $\hbar\to0$.  Using Eq.~(\ref{eq9}), we find an approximate solution, starting from the initial ground-state wavefunction $(\alpha,\beta)|_{t=t^{\rm{in}}}=(1,0)$, as
\begin{align}
	\psi = \psi_{-} - \sum_{n}  (-1)^{n} {\rm{e}}^{-\frac{\sigma_n}{\hbar}}  \Theta(t-t^{\rm{cr}}_{n}) \psi_{+} . \label{eq13}
\end{align}

The Stokes constant $\beta$ (\ref{eq9}) [i.e., the coefficient in front of $\psi_+$ in Eq.~(\ref{eq13})] gives the probability amplitude from the initial ground state to an excited state with the instantaneous basis (\ref{eq-4}).  Thus, $\beta\neq0$ means that electron-hole pair production occurs (see Fig.~\ref{fig-2}), and $|\beta|^2$ gives the pair production number.  The typical magnitude of the production number is determined by the exponential factor ${\rm{e}}^{-\frac{1}{\hbar}{\rm{Re}}\,\sigma_n}$.  Note that Eq.~\eqref{eq13} takes the form of superposition of the wavefunctions describing the pair production at different times.  The phase factor ${\rm{e}}^{-\frac{\rm{i}}{\hbar}{\rm{Im}}\,\sigma_n}$, related to the so-called St\"{u}ckelberg phase, is thus responsible for quantum interference effect~\cite{Dumlu:2010ua, Dumlu:2011rr, Shevchenko_2010, stuckelberg1933theorie}: When pair production occurs repetitively, the St\"{u}ckelberg phase, i.e., the relative phase between each wavefunction
\begin{align}
	\theta_{n,n'} \equiv \frac{2}{\hbar}\int_{t^{\rm{cr}}_{n'}}^{t^{\rm{cr}}_n}{\rm }{\rm{d}}t\,\epsilon = -\frac{1}{\hbar} ({\rm{Im}}\,\sigma_{n} - {\rm{Im}}\,\sigma_{n'}) \neq 0 ,
\end{align}
gives rise to destructive and constructive interferences, which suppress and enhance the production, respectively.  Note that our Stokes constants $\alpha$ and $\beta$ (\ref{eq9}) neglect pair annihilation processes followed after the production, which are higher order effects ${\mathcal{O}}({\rm{e}}^{-\frac{2}{\hbar}{\rm{Re}}\,\sigma_n})$.

Within the lowest-order WKB approximation (\ref{eq-4}), the production number $|\beta|^2$ shows stepwise time-dependence, meaning that each production process is treated as an instantaneous one that occurs exactly at $t=t^{\rm{cr}}_n$.  In reality, the stepwise evolution is an approximation, and pair production takes finite time $\delta{t}\neq0$.  Under an electric field, for a pair production to occur via quantum tunneling, an electron in the valence-band needs to tunnel a distance $\sim\Delta/eE_0$ to the conduction-band, and hence $\delta{t}\sim\gamma\Omega^{-1}$ with the Keldysh parameter $\gamma$.  Thus, the WKB approximation would not work for very high harmonics $N\equiv\omega/\Omega\gtrsim(2\pi/\delta{t})/\Omega\sim2\pi\gamma^{-1}$ or when $\gamma\gtrsim2\pi/N$.  This also implies that the WKB approximation cannot describe perturbative excitation processes, which dominate for large $\gamma$~\cite{Keldysh:1965ojf,Brezin:1970xf,Popov:1971ff,Taya:2014taa,Taya:2020dco}.

{\it WKB result for electric current.}\ We compute the electric current $J\equiv\braket{-\delta{H}/\delta{A}}=e\psi^\dagger\sigma^1\psi$ at each momentum $k$.  The WKB approximation (\ref{eq13}) gives $J$ as a sum of the valence-, intra-, and inter-band contributions, $J=J_{\rm{val}}+J_{\rm{intra}}+J_{\rm{inter}}$, where
\begin{align} 
	J_{\rm{val}} 
		&\!\equiv e\psi^\dagger_- \sigma^1 \psi_-
		= -ev  ,\label{eq14} \\
	J_{\rm{intra}} 
		&\!\equiv\! \sum_{\pm} \!|\beta|^2 \!(\pm e) \psi^\dagger_\pm \sigma^1 \!\psi_\pm 
		\!=\! 2ev \!\left| \sum_{n} \!(-1)^{n} \!{\rm{e}}^{-\!\frac{ \sigma_n}{\hbar} } \!\Theta(t-t^{\rm{cr}}_{n}) \right|^2 \!\!, \nonumber\\
	J_{\rm{inter}} 
		&\!\equiv 2e\,{\rm{Re}}\left( \alpha \beta^* \psi^\dagger_+ \sigma^1 \psi_- \right) \nonumber\\
		&=\! e\frac{\Delta}{\epsilon} \!\sum_{n} (-1)^{n} {\rm{e}}^{-\frac{1}{\hbar} {\rm{Re}}\,\sigma_n } \Theta(t-t^{\rm{cr}}_{n}) \cos \!\left( \frac{2}{\hbar} \int^t_{t^{\rm{cr}}_{n}} \!{\rm{d}}t\,\epsilon \right) , \nonumber
\end{align}
with velocity $v\equiv(k-eA)/\epsilon$.  Being proportional to $v$, $J_{\rm{val}}$ ($J_{\rm{intra}}$) is accompanied by valence-band electrons (electron-hole pairs) accelerated along the band(s).  $J_{\rm{inter}}$ originates from interference between conduction- and valence-band electrons (or dipole \cite{Tanji:2008ku}), as is evident from the wavefunction overlap ${\rm{Re}}(\alpha\beta^* \psi^\dagger_+\sigma^1\psi_-)$.  Since the actual observable is the difference from the ground state, we subtract $J_{\rm{val}}$ and focus on 
\begin{align}
	J_{\rm{obs}} \equiv J - J_{\rm{val}} = J_{\rm{intra}} + J_{\rm{inter}} . 
\end{align}
This is a standard subtraction scheme widely used in quantum-field theory in external fields~\cite{Birrell:1982ix}. Also, in the condensed-matter context, the subtracted $J_{\rm val}$ amounts to be zero when summed over the entire Brillouin zone. Note that the observable currents, $J_{\rm{intra}}$ and $J_{\rm{inter}}$, reproduce the phenomenological expressions used in previous studies on semi-conductor HHG~\cite{Vampa2014, Vampa2015}.

According to Eq.~(\ref{eq14}), the nonlinear response to the applied field arises not only from the pair production factor ${\rm{e}}^{-\frac{\sigma_n}{\hbar}}$ but also from the velocity $v$ in $J_{\rm intra}$.  Whereas the nonlinearity of $v$ is absent for the quadratic dispersion, it can be significant for the linear one in massless Dirac systems~\cite{Mikhailov_2007,Cheng2020,Kovalev2020}.  The strong nonlinearity survives to some extent even in our massive case.

\begin{figure}[t]
\begin{center}
\hspace*{-7mm}
\includegraphics[clip, width=0.4\textwidth]{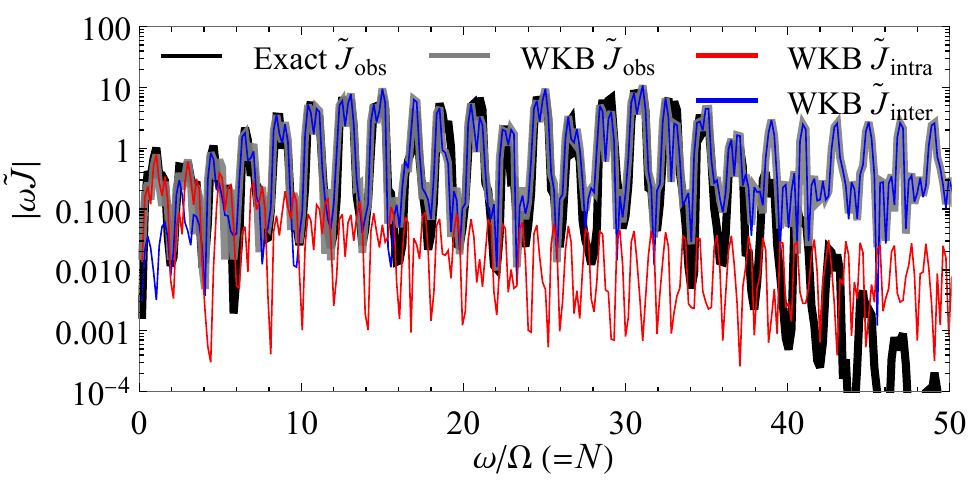}
\vspace*{-7mm}
\end{center}
\caption{\label{fig-3} HHG spectrum for the oscillating field (\ref{eq-9}), with $\Omega/(\Delta/2)=1/4,\,eE_0/(\Delta/2)^2=1$ (i.e., $\gamma=1/4$) and ${\Omega}t_{\rm{in}}=-17\pi/3,\,T_w=2|t_{\rm{in}}|$.  The parameter set corresponds to, e.g., $\Omega/2\pi=1\,\mathrm{THz}$ and $E_0=4.2\,\mathrm{kV/cm}$ for a Dirac material with Fermi velocity $v_{\rm F}=10^6\,\mathrm{m/s}$ and mass $\Delta=33\,\mathrm{meV}$.  
}\vspace*{-5mm}
\end{figure}

{\it HHG by ac electric field.}\ As a demonstration, we consider a monochromatic ac field, 
\begin{align}
	eA = - \frac{eE_0}{\Omega} \sin(\Omega t) , \label{eq-9} 
\end{align}
and take $k\to0$, as pair production at large $k$ may be energetically disfavored (see Ref.~\cite{Yue2021} for situations where various $k$'s become important).  For the field (\ref{eq-9}), the wavefunction encounters two crossings in a cycle $t_n^{{\rm{cr}}}\in\frac{\pi}{\Omega}{\mathbb{Z}}$, i.e., pairs are produced twice in a cycle.  This is similar to the three-step process in the gas HHG~\cite{PhysRevLett.71.1994,PhysRevA.49.2117}.  Although our WKB approach also applies to other $k$'s, their analytical expressions become more complex. We leave, for future work, this generalization and summing the results over $k$'s. In exchange for restricting ourselves to $k\to0$, we will obtain analytical closed formulas, which would at least qualitatively capture the HHG in the massive Dirac electrons.

We compute the Fourier spectrum of the observable $\tilde{J}_{\rm{obs}}\equiv\int^{+\infty}_{-\infty}{\rm{d}}t\,{\rm{e}}^{+{\rm{i}}\omega{t}}WJ_{\rm{obs}}=\tilde{J}_{\rm{intra}}+\tilde{J}_{\rm{inter}}$, where we insert a window function $W$ with width $T_w$ to avoid contaminations due to the finiteness of fields/measurements~\cite{Wu2015,PhysRevResearch.2.032015}.  Here, we chose the Hann window $W\equiv\Theta(t-t_{\rm{in}})\Theta(T_w+t_{\rm{in}}-t)\sin^2(\pi(t-t_{\rm{in}})/T_w)$ and confirmed that the results are insensitive to the choice of $W$.  One can obtain a closed analytical expression for $J_{\rm{obs}}$ under Eq.~(\ref{eq-9}), from which one can compute the spectrum $\tilde{J}_{\rm{obs}}$ numerically or even analytically under certain approximations (see Appendix~\ref{app1A}).  

The HHG spectrum obtained from the WKB approximation (\ref{eq14}) agrees well for small $\gamma$ with the exact one obtained by numerically solving TDSE; see Fig.~\ref{fig-3}.  Due to the limitation of the WKB approximation, the discrepancy for very high harmonics appears at $N\gtrsim35$ in Fig.~\ref{fig-3}, which is consistent with our estimate $\sim{2}\pi\gamma^{-1}$.  

Our WKB result confirms that the interband contribution dominates over the intraband one, except for the low harmonics ($N\lesssim3$ in Fig.~\ref{fig-3})~\cite{Vampa2014, Vampa2015}.  This means that the interband contribution is the origin of the plateau structure in the HHG spectrum~\cite{schubert2014sub}.  One may estimate the location of the cutoff by using the saddle-point method~\cite{Vampa2014, Vampa2015,PhysRevA.49.2117} when evaluating the Fourier integral of the WKB expressions (\ref{eq14}).  The stationary conditions for the intra- and inter-band currents are given by $\omega=0$ and $2\epsilon-\hbar|\omega|=0$, respectively.  The former condition confirms why the intraband current contributes only to the low harmonics.  The latter condition can have real solutions when $\Delta/\hbar\Omega\lesssim{N}\lesssim\sqrt{\Delta^2+4(|k|+eE_0/\Omega)^2}/\hbar \Omega\sim2(|k|+eE_0/\Omega)/\hbar\Omega$.  If the stationary points are imaginary, the saddle-point action acquires positive real part and then the spectrum $\tilde{J}_{\rm{inter}}$ is suppressed exponentially.  Therefore, it is sufficient to pick up the contribution from the real solution only, which yields that the upper cutoff of the HHG spectra for $k\to0$ is given by 
\begin{align}
	N_{\rm cut}\sim\frac{2eE_0}{\hbar\Omega^2}
\end{align}
($\sim32$ in  Fig.~\ref{fig-3}).  The linear dependence on $E_0$ is consistent with semiconductor experiments~\cite{ghimire2011observation}, and the frequency dependence $\Omega^{-2}$ is our prediction in the massive Dirac electrons, which is worth testing in experiments (see Ref.~\cite{Ikeda2018b} for a similar prediction in a charge-density-wave material).  Note that the lower cutoff $N\sim\Delta/\hbar\Omega$ ($\sim8$ in Fig.~\ref{fig-3}) for the interband current is also consistent with Fig.~\ref{fig-3}.

\begin{figure*}[t]
\begin{flushleft}
\mbox{(a)}
\end{flushleft}
\vspace*{-10mm}
\begin{center}
\hspace*{-7mm}
\includegraphics[clip, height=0.155\textwidth]{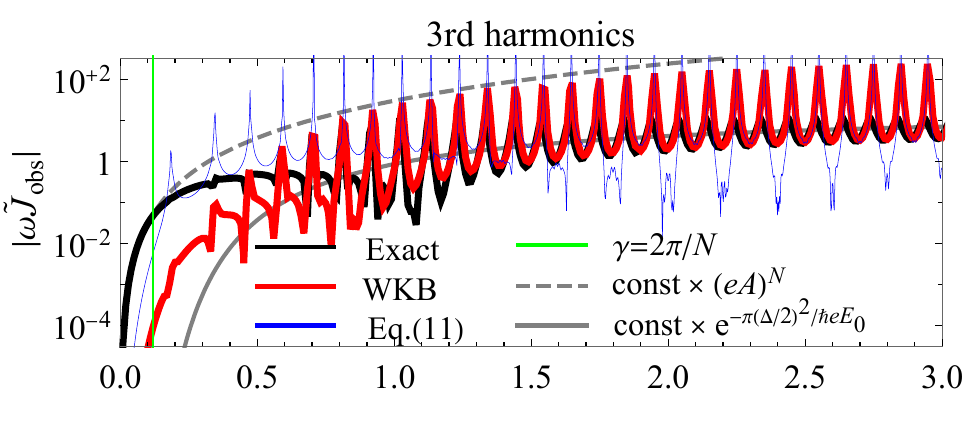}\hspace*{-0mm}
\includegraphics[clip, height=0.156\textwidth]{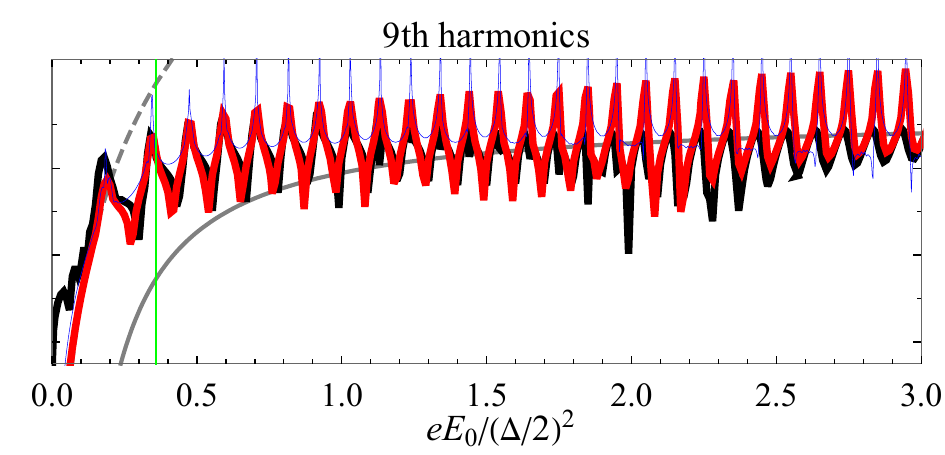}\hspace*{-0mm}
\includegraphics[clip, height=0.155\textwidth]{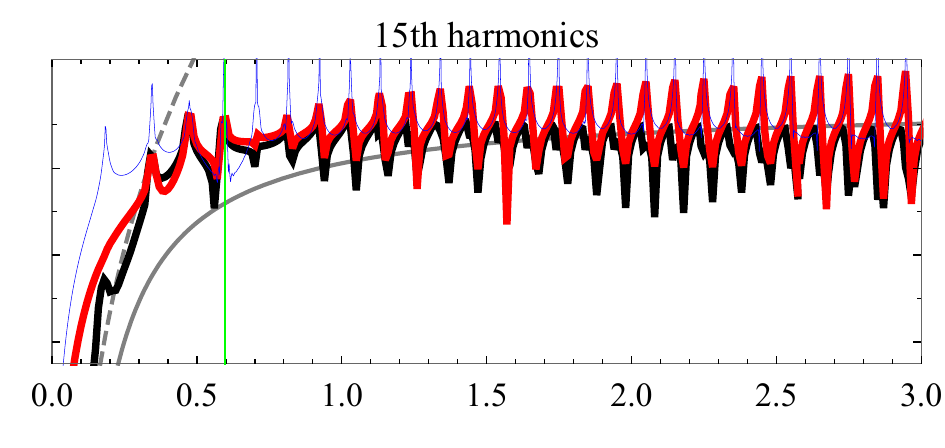}
\end{center}
\vspace*{-10.5mm}
\begin{flushleft}
\mbox{(b)}
\end{flushleft}
\vspace*{-7mm}
\begin{center}
\hspace*{-7mm}
\includegraphics[clip, height=0.139\textwidth]{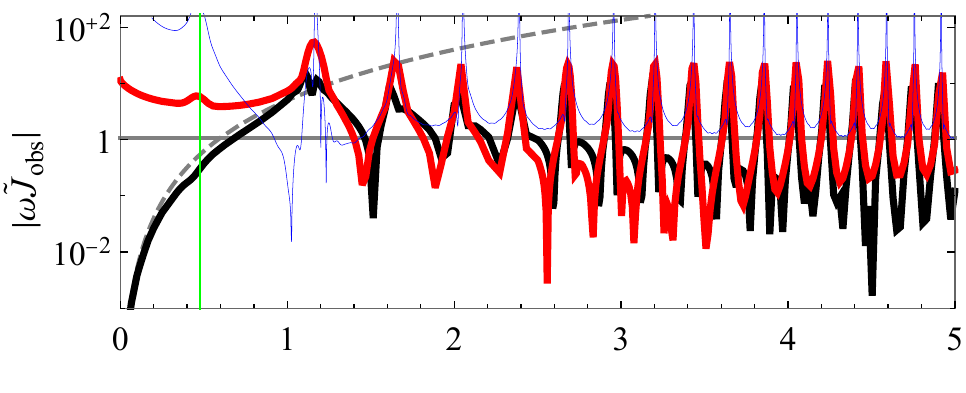}\hspace*{1.9mm}
\includegraphics[clip, height=0.135\textwidth]{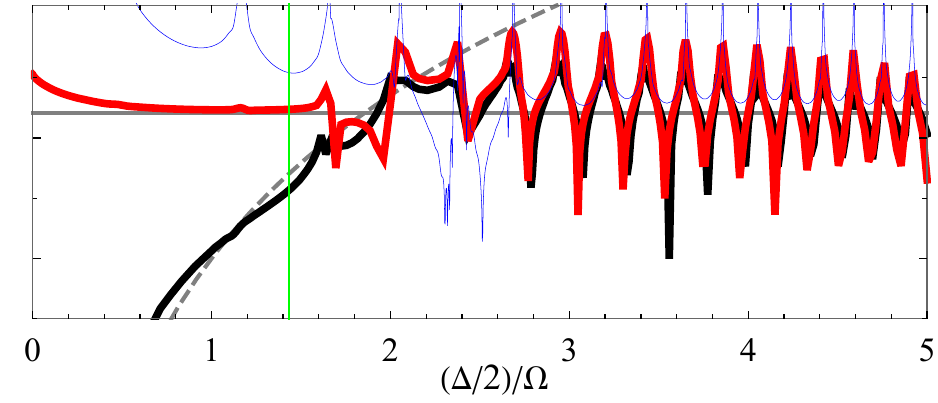}\hspace*{1.9mm}
\includegraphics[clip, height=0.136\textwidth]{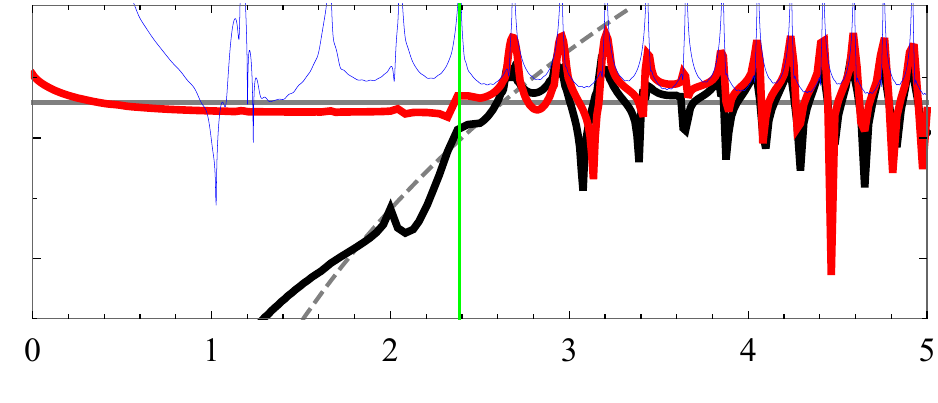}
\end{center}
\vspace*{-7mm}
\caption{\label{fig-4} The magnitude of $N=3$ (left), 9 (middle), and 15 (right) harmonic peaks plotted against amplitude $eE_0$ with fixed frequency $\Omega/(\Delta/2)=1/4$ (a) and inverse frequency $\Omega^{-1}$ with fixed amplitude $eE_0/(\Delta/2)^2=1$ (b).  The other parameters are chosen as ${\Omega}t_{\rm{in}}=-17\pi/3,\,T_w=2|t_{\rm{in}}|$.  }\vspace*{-5mm}
\end{figure*}

To further investigate the HHG spectrum, we plot the $eE_0$- and $\Omega^{-1}$-dependence of each harmonic intensity in Fig.~\ref{fig-4}.  The WKB result reproduces the exact one in the nonperturbative high-intensity and low-frequency regime $\gamma\lesssim2\pi/N$, while it fails for the perturbative low-intensity and high-frequency one $\gamma\gtrsim2\pi/N$.  In the perturbative regime, the $N$-th harmonic intensity shows power-dependence on the potential $(eA)^N\propto(eE_0/\Omega)^N$.  This indicates that a perturbative $N$-photon process dominates, which cannot be captured by the WKB approximation.  On the other hand, the harmonic intensities saturate (with oscillations) in the nonperturbative regime.  The typical magnitude of the currents asymptote $|\tilde{J}_{\rm{intra}}|\to|\beta|^2\propto{\rm{e}}^{-\pi\frac{(\Delta/2)^2}{eE_0}}$ and $|\tilde{J}_{\rm{inter}}|\to|\beta|\propto{\rm{e}}^{-\pi\frac{(\Delta/2)^2}{2eE_0}}$, which are independent of $\Omega$ and $N$.  This explains why the harmonic intensities have weak $N$-dependence in the nonperturbative regime and the plateau appears in Fig.~\ref{fig-3}.  Notice that the nonperturbative dependence on $eE_0$ is the manifestation that the production is driven by tunneling and agrees with the tunneling production formula~\cite{landau1937theorie,zener1932non,stuckelberg1933theorie,1932NCim....9...43M,Sauter:1931zz,Heisenberg:1935qt,Schwinger:1951nm}.

The oscillating behavior in Fig.~\ref{fig-4} is caused by the quantum interference.  When the interference becomes destructive (constructive), the production and associated HHG are suppressed (enhanced).  For our field (\ref{eq-9}) with the limit $k\to 0$, the St\"{u}ckelberg phase can be decomposed as $\theta_{n,n'}=(n-n')\theta$, where $\theta\equiv\frac{2}{\hbar}\int_{t^{\rm{cr}}_{n}}^{t^{\rm{cr}}_{n+1}}{\rm{d}}t\,\epsilon=\frac{2}{\hbar}\int^{\pi/\Omega}_{0}{\rm{d}}t\,\epsilon$ is the phase for two successive production and is independent of $n$.  Then, Eq.~(\ref{eq9}) indicates that the most destructive (constructive) interference occurs when $\theta$ matches even- (odd-) integer multiples of $\pi$.  As shown analytically below, $\theta$ is roughly proportional to $eE_0/\Omega^2$, and hence the harmonic intensities oscillate with $eE_0$ and $\Omega$.  This St\"{u}ckelberg-phase mechanism for the oscillation is analogous to that in the gas HHG~\cite{Toma1999,Popruzhenko2002} and would also apply to the recent semiconductor HHG experiment~\cite{xia2020highharmonic}.
Note that we focus on $k\to0$ although the physically-observed current are obtained in principle by collecting contributions from different $k$'s over the Brillouin zone. Since the St\"{u}ckelberg phase depends smoothly on $k$, the sum over $k$ would smoothen the oscillating behavior in Fig.~\ref{fig-4}. As noted above, we leave it for future work to extend our analytical WKB theory for different $k$'s.

{\it Analytical formula in the low-frequency limit.}\ To get deeper insights, we analytically carry out the Fourier integration for the WKB result in the limit of $\gamma,k\to0$ (see Appendix~\ref{app1A}):
\begin{widetext}
\begin{align}
	\tilde{J}_{\rm{intra}}
		&\sim {\rm{e}}^{-\frac{2\,{\rm{Re}}\,\sigma}{\hbar} } \frac{-{\rm{i}}}{\pi \cos^2\frac{\theta}{2}}\sum_{n=-\infty}^\infty \left[ \sum_{\pm} {\rm{e}}^{+{\rm{i}} \pi \left( \left\lceil \frac{t_{\rm{in}}}{\pi/\Omega}   \right\rceil -\frac{1}{2} \right) \left( \mp \frac{\theta}{\pi}  -1 \right)} \frac{ \sin\frac{\theta}{2}}{4\left(n-\frac{\theta}{2\pi}\right)}  \tilde{W} \left(  \omega -\Omega \mp 2\left(n-\frac{\theta}{2\pi}\right) \Omega \right) -\frac{\tilde{W} \left( \omega - (2n-1)\Omega \right) }{2n-1} \right]  ,  \nonumber\\
	\tilde{J}_{\rm{inter}}
		&\sim {\rm{e}}^{-\frac{{\rm{Re}}\,\sigma}{\hbar} } \frac{-{\rm{i}}(-1)^{ \left\lceil \frac{t_{\rm{in}}}{\pi/\Omega} \right\rceil } \gamma}{4\pi\cos\frac{\theta}{2}} \sum_{n=-\infty}^{+\infty} \sum_\pm  \left[ \left( {\rm ln}\,\gamma^2 + 2H_{\pm n-1/2} \right) {\rm{e}}^{ +{\rm{i}} \pi \left( \left\lceil \frac{t_{\rm{in}}}{\pi/\Omega}   \right\rceil  -\frac{1}{2} \right) \left(\mp \frac{\theta}{\pi} - 1\right) } \tilde{W} \left( \omega \mp 2\left(n-\frac{\theta}{2\pi}\right)\Omega \right)  \right] \nonumber\\
			&\quad \left. + \left( {\rm ln}\,(4\gamma^2) - \left( (-1)^n \cos\frac{\theta}{2} - 1  \right) H_{\frac{n}{2}\pm \frac{\theta}{4\pi}-1} + \left(  (-1)^n \cos\frac{\theta}{2} + 1  \right) H_{\frac{n}{2} \pm \frac{\theta}{4\pi}-\frac{1}{2}} \right)  \tilde{W} \left(  \omega - (2n -1) \Omega \right) \right] ,   \label{eq_10} 
\end{align} 
\end{widetext}
where $\tilde{W}$ is the Fourier transform of the window function (i.e., a regularized delta function), $H_n$ is the harmonic number, ${\rm{Re}}\,\sigma=\pi\frac{(\Delta/2)^2}{eE_0}[1+{\mathcal{O}}(\gamma^2)]$, and $\theta=\frac{4eE_0}{\hbar\Omega^2}[1+{\mathcal{O}}(\gamma^2)]$.

The analytic formula (\ref{eq_10}) captures the important features of the exact numerical results for small $\gamma$; see Fig.~\ref{fig-4}.  The saturation behavior is determined by the overall factor ${\rm{e}}^{-\frac{{\rm{Re}}\,\sigma}{\hbar}}$, reflecting the abundance of pair production due to tunneling.  The oscillating behavior derives from $\cos\frac{\theta}{2}$ in the denominators.  Whenever $\theta$ hits an odd-integer multiple of $\pi$, each harmonic intensity is maximized, as  anticipated from the quantum-interference argument.  Incidentally, the second terms in the square brackets give odd-order harmonics, while the first ones give non-integer split peaks around the integer harmonics with $\delta{N}=\pm\theta/\pi$.  This splitting is a transient effect, as is evident from the $t_{\rm{in}}$-dependence, and is tunable by changing the St\"{u}ckelberg phase $\theta$, which can be varied, e.g., with the carrier-envelop phase~\cite{Schmid2021}.

{\it Summary.}\ We have studied HHG in a massive Dirac system based on the lowest-order WKB approximation, including Stokes phenomena.  We have shown that the WKB approximation provides a powerful analytical framework to study HHG in the nonperturbative low-frequency and high-intensity regime and well reproduces the exact results of TDSE.  Our results imply that the repetitive dynamics of production and acceleration of electron-hole pairs and quantum interference due to the St\"{u}ckelberg phase are the essence of HHG.  We have also predicted some characteristic features of HHG in the deep nonperturbative regime, such as the scaling of the cutoff $N_{\rm{cut}}\propto{eE_0}/\Omega^2$, the oscillation of harmonic intensities with a period determined by the St\"{u}ckelberg phase $\theta{\propto}eE_0/\Omega^2$, and the non-integer splittings of harmonic peaks $\delta{N}\propto\theta$ as a transient effect.  Our WKB approach applies to various media such as Dirac/Weyl materials (including QED), topological insulators, semiconductors, and graphene, paving the way toward a universal understanding of HHG beyond gases.   

{\it Acknowledgments.}\ The authors thank the Yukawa Institute for Theoretical Physics at Kyoto University, where this work was initiated during YITP-T-20-05 “The Schwinger Effect and Strong-Field Physics Workshop,” and are supported by Non-Equilibrium Working group (NEW) at RIKEN Interdisciplinary Theoretical and Mathematical Sciences Program (iTHEMS).  H.T. thanks Antonino~Di~Piazza for useful discussions.  M.H. was supported by the US Department of Energy, Office of Science, Office of Nuclear Physics under Award No. DE-FG0201ER41195.  T.N.I. was supported by JSPS KAKENHI Grants No.~JP18K13495 and No.~JP21K13852.

\appendix

\section*{High-harmonic spectrum for $k=0$} \label{app1A}

We explain details of the computation of the Fourier spectrum $\tilde{J}_{\rm obs}$ for the oscillating electric field $eA(t)=-\frac{eE_0}{\Omega} \sin (\Omega t)$ in the limit of $k = 0$.  

\subsection{Computation of current $J_{\rm obs}(t)$ in the coordinate space}

We begin by deriving an explicit expression for the current in the coordinate space $J_{\rm obs}(t)$.  For this purpose, it is convenient to introduce two dimensionless parameters, 
\begin{align}
	\gamma \equiv \frac{\Omega (\Delta/2)}{eE_0},\ 
	\nu \equiv \frac{ eE_0 }{\Omega^2} .  
\end{align}
Any physical observables, including $J_{\rm obs}$, can be expressed solely in terms of those two dimensionless parameters (plus the time variable $t$, which we always scale with the frequency $\Omega$ as $\Omega t$) because the system has in total three dimensionful parameters $eE_0, \Omega$ and $\Delta$, out of which one can construct only two dimensionless quantities.  In terms of $\gamma$ and $\nu$, one can show that the turning points $z^{\rm tp}_n$'s are located at
\begin{subequations}  
\begin{align}
	z^{\rm tp}_n = -\frac{\rm i}{\Omega} {\rm ln}\left[ \gamma + \sqrt{1+\gamma^2} \right] + (n+n_{\rm in}-1) \frac{\pi}{\Omega} ,
\end{align}
\end{subequations}
where $n=1,2,\cdots \in {\mathbb N}$ and 
\begin{align}
	n_{\rm in} 
		\equiv \left\lceil \frac{t_{\rm in}}{\pi/\Omega} \right\rceil
		= \Bigl\{ m \in{\mathbb Z}\ \Bigl|\  m-1<\frac{t_{\rm in}}{\pi/\Omega}\leq m  \Bigl\} .
\end{align}
The associated crossings $t^{\rm cr}_{n}$'s and integrals $\sigma_n$ read 
\begin{align}
	t^{\rm cr}_{n}  
		= {\rm Re}\,z^{\rm tp}_{n} 
		= (n+n_{\rm in}-1) \frac{\pi}{\Omega} ,
\end{align}
and 
\begin{align}
	\sigma_n
		= S - {\rm i}\rho\left( - (n+n_{\rm in}-1) \frac{\pi}{\Omega} \right) ,
\end{align}
where
\begin{subequations}
\begin{align}
	S	&\equiv 2{\rm i}\int_{0}^{-\frac{\rm i}{\Omega} {\rm ln}\left[ \gamma + \sqrt{1+\gamma^2} \right]} {\rm d}z\,\epsilon(z) \nonumber\\
		&= 2\nu \left[ (1+\gamma^2){\rm K}(-\gamma^2)-{\rm E}(-\gamma^2) \right] \nonumber\\
		&= {\rm Re}\,\sigma_n, \\
	\rho(t) 
		&\equiv 2  \int_{0}^{t} {\rm d}z\,\epsilon(z) \nonumber\\
		&= 2\nu\gamma\,{\rm E}(\Omega t; -\gamma^{-2}) ,    \label{eqs6b}
\end{align}
\end{subequations}
with ${\rm K}(k)$ being the complete elliptic integral of the first kind and ${\rm E}(k)$ and ${\rm E}(\varphi; k)$ being the complete and incomplete elliptic integrals of the second kind, respectively.  Note that $\rho (t + n\pi/\Omega) = \rho(t) + n \rho(\pi/\Omega)$ and that the St\"{u}ckelberg phase for the two successive production times $\theta$ can be expressed in terms of $\rho$ as
\begin{align}
	\theta	\equiv \frac{2}{\hbar} \int_{0}^{\pi/\Omega} {\rm d}t\,\epsilon(t) 
		= \frac{1}{\hbar} \rho(\pi/\Omega)
		= \frac{4}{\hbar} \nu \gamma \, {\rm E}(-\gamma^{-2}) .   \label{eq79}
\end{align}
Note that $\theta$ is related to (a relativistic generalization of) the pondermotive energy (i.e., the average energy of an electron in a cycle) $U$ as $U \equiv \int^{2\pi/\Omega}_0 {\rm d}t\,\epsilon / (2\pi/\Omega) = \theta \hbar\Omega/2\pi$.

Substituting the above into the WKB expression for the current $J_{\rm obs}(t) = J_{\rm intra}(t)+J_{\rm intra}(t)$, one obtains
\begin{subequations}
\label{eq82} 
\begin{align}
	J_{\rm intra} (t)
		&= 2 ev(t) {\rm e}^{-\frac{2S}{\hbar}  } \left|  \sum_{n=n_{\rm in}}^{\infty} (-1)^{n} {\rm e}^{+{\rm i} n \theta} \Theta(t-n \pi/\Omega) \right|^2 , \\
	J_{\rm inter} (t)
		&= 2e m(t) (-1)^{-n_{\rm in}+1} {\rm e}^{-\frac{S}{\hbar} } \nonumber\\
		&\quad \times \sum_{n=n_{\rm in}}^{\infty} (-1)^{n}\Theta(t-n \pi/\Omega) \cos \left( \frac{\rho(t-n \pi /\Omega)  }{\hbar} \right) .  
\end{align}
\end{subequations}
where
\begin{align}
	v(t)	
		&= \frac{-eA(t)}{\epsilon(t)} 
		= \frac{ \gamma^{-1} \sin(\Omega t)}{\sqrt{ 1 + \gamma^{-2} \sin^2(\Omega t)}} ,   \nonumber\\
	m(t) 
		&\equiv \frac{\Delta/2}{\epsilon(t)}
		= \frac{1}{\sqrt{ 1 + \gamma^{-2} \sin^2(\Omega t)}} .  
\end{align}
The factor ${\rm e}^{-\frac{S}{\hbar} }$ controls the abundance of the pair production, which just changes the overall size of the currents and does not affect the peak structure of the Fourier spectrum.  The peak structure is determined by the summation factors $ \sum_{n=n_{\rm in}}^{\infty} (-1)^{n} {\rm e}^{+{\rm i} n \theta} \Theta(t-n \pi/\Omega)$ and $\sum_{n=n_{\rm in}}^{\infty} (-1)^{n}\Theta(t-n \pi/\Omega) \cos \left( \frac{\rho(t-n \pi /\Omega)  }{\hbar} \right) $.  The step functions describe when the pair production processes occur, while the phase factors are responsible for the quantum interference.  Because of the periodicity of the applied field, the pair-production points are periodic with the period $\pi/\Omega$, which essentially results in that the Fourier spectrum is periodically peaked with frequency $2\pi/(\pi/\Omega) = 2\Omega$.

\subsection{Computation of the Fourier spectrum $\tilde{J}^\infty_{\rm obs}(\omega)$ without convoluting a window function}

We evaluate the Fourier spectrum for the current $J_{\rm obs} = J_{\rm intra} + J_{\rm inter}$, without convoluting any window functions, i.e., we evaluate
\begin{align}
	\tilde{J}^\infty_{\rm obs} (\omega)
		= \int^{+\infty}_{-\infty} {\rm d}t\,{\rm e}^{+{\rm i}\omega t} J_{\rm obs}(t) 
		= \tilde{J}^\infty_{\rm intra}(\omega) + \tilde{J}^\infty_{\rm inter}(\omega) .  
\end{align}
Note that $\tilde{J}^\infty_{\rm obs}(\omega)  = \lim_{T_w\to \infty} \tilde{J}_{\rm obs}(\omega) $

We first evaluate $\tilde{J}^\infty_{\rm intra}$.  One may express $\tilde{J}^\infty_{\rm intra}$ as a convolution integral with $\tilde{v}$ as
\begin{widetext}
\begin{align}
	\tilde{J}^\infty_{\rm intra}(\omega) 
		&= \int^{+\infty}_{-\infty} {\rm d}t\,{\rm e}^{+{\rm i}\omega t} J_{\rm intra}(t) \nonumber\\
		&= \int^{+\infty}_{-\infty} \frac{{\rm d}\omega'}{2\pi} e\tilde{v}(\omega-\omega') \underbrace{ 2 \int^{+\infty}_{-\infty} {\rm d}t\,{\rm e}^{+{\rm i}\omega' t} {\rm e}^{-\frac{2S}{\hbar}  }  \sum_{n=n_{\rm in}}^{\infty} \frac{ \cos\left( \left( 2(n-n_{\rm in}) + 1  \right)\frac{\theta}{2} \right) }{ \cos \frac{\theta}{2}}  (-1)^{n_{\rm in}+n} \Theta(t-n \pi/\Omega) }_{\equiv f_{\rm intra}(\omega')} .  \label{eqs10} 
\end{align}
\end{widetext}
Since $v$ is an odd function and is periodic with the period $2\pi/\Omega$, it is legitimate to expand $v$ with a Fourier sine series as
\begin{align}
	v(t) = \sum_{n=1}^\infty s_n \sin(n \Omega t) , 
\end{align}
where
\begin{align}
	s_n 
		&\equiv \frac{\Omega}{\pi} \int^{+\pi/\Omega}_{-\pi/\Omega} {\rm d}t\,\sin(n \Omega t) v(t) \nonumber\\
		&= \frac{1}{\pi} \int^{+\pi}_{-\pi} {\rm d}x \frac{ \gamma^{-1} \sin(x) \sin(nx) }{\sqrt{1+\gamma^{-2}\sin^2 (x)  }} .
\end{align}
After some calculations, one can show that 
\begin{align}
	s_n &= \left\{ \begin{array}{ll}
				0 & (n:\ {\rm even}) \\[8pt]
				(-1)^{\left\lfloor \frac{n}{2} \right\rfloor} \gamma^{-1} {}_3\tilde{F}_2\left( \begin{array}{c} \frac{1}{2},1,\frac{3}{2} \\ \frac{3-n}{2}, \frac{3+n}{2} \end{array}; -\gamma^{-2} \right) & (n:\ {\rm odd}) 
			 \end{array} \right. , \label{eq104}
\end{align}
where ${}_p\tilde{F}_q$ is the regularized hypergeometric function.  Therefore, 
\begin{align}
	\tilde{v}(\omega) 
		&= \sum_{n=1}^\infty s_n \int^{+\infty}_{-\infty}{\rm d}t\,{\rm e}^{+{\rm i}\omega t} \sin(n \Omega t) \nonumber\\
		&= -{\rm i}\pi \sum_{n=-\infty}^{+\infty} (-1)^n \gamma^{-1} {}_3\tilde{F}_2\left(\begin{array}{c} \frac{1}{2},1,\frac{3}{2} \\ 1+n, 2-n \end{array}; -\gamma^{-2} \right) \nonumber\\
		&\quad \times \delta(\omega - (2n-1)\Omega ) , \label{eq106}
\end{align}
which has odd harmonics only.  Note that $\lim_{\gamma \to \infty} {}_3\tilde{F}_2\left(\begin{array}{c} \frac{1}{2},1,\frac{3}{2} \\ 1+n, 2-n \end{array}; -\gamma^{-2} \right) \delta(\omega - (2n-1)\Omega ) \propto (\delta_{n,0}+\delta_{n,1})$, i.e., the velocity $\tilde{v}$ only gives the lowest harmonics $|\omega| = \Omega$ in the nonrelativistic limit $\Delta \to \infty$ where $\gamma \to \infty$.  Conversely, $\tilde{v}$ contributes to harmonics higher than the lowest one because of the relativistic nature of $v$ when $\gamma < \infty$.  Our remaining task is to evaluate $f_{\rm intra}(\omega)$.  By adding an infinitesimal imaginary part to the frequency $\omega \to \omega + {\rm i}0^+$ so that the Fourier integral converges and is well-defined, we find
\begin{widetext}
\begin{align}
	f_{\rm intra}(\omega)
		= {\rm i}\,{\rm e}^{-\frac{2S}{\hbar} } \frac{{\rm e}^{+{\rm i} \pi \left( n_{\rm in}-\frac{1}{2} \right) \frac{\omega}{\Omega} }  }{\omega + {\rm i}0^+ } \frac{ 1}{ \cos \frac{\theta}{2} } \frac{1}{2} \left[ \frac{1}{\cos \left( \pi \left( \frac{\omega+{\rm i}0^+}{2\Omega} + \frac{\theta}{2\pi} \right) \right)} + \frac{1}{\cos \left( \pi \left( \frac{\omega+{\rm i}0^+}{2\Omega}  - \frac{\theta}{2\pi}  \right)\right)}  \right]  , \label{eqs16} 
\end{align}
which is peaked at $\omega/\Omega = 0, (2n-1) \mp \frac{\theta}{\pi}$.  Indeed,
\begin{align}
	f_{\rm intra}(\omega)
		&= \frac{1}{\Omega} \frac{{\rm e}^{-\frac{2S}{\hbar} }}{\cos\frac{\theta}{2}} \left[ \frac{ \pi}{\cos\frac{\theta}{2}}\delta\left( \frac{\omega}{\Omega} \right) 
			 + \frac{{\rm e}^{+{\rm i} \pi \left( n_{\rm in}-\frac{1}{2} \right)\frac{\omega}{\Omega}}}{\omega/\Omega} \sum_\pm \sum_{n=-\infty}^{+\infty} (-1)^n  \delta \left( \frac{\omega}{\Omega} \pm \frac{\theta}{\pi} - (2n-1)  \right) \right] + ({\rm finite}) ,   \label{eq102}
\end{align}
where we used $\frac{1}{\omega + {\rm i}0^+} = -{\rm i}\pi \delta(\omega) + ({\rm finite})$ and $\frac{1}{\cos(\omega + {\rm i}0^+ )} = -{\rm i}\pi \sum_{n=-\infty}^{+\infty} (-1)^n \delta \left( \omega - \frac{(2n-1)\pi}{2} \right) + ({\rm finite})$.  The first term is the source of odd higher harmonics, after convoluting with $\tilde{v}$, and the second one gives rise to HHG with splittings, as we show just below.  Substituting Eqs.~(\ref{eq106}) and (\ref{eqs16}) into Eq.~(\ref{eqs10}), we arrive at 
\begin{align}
	\tilde{J}^\infty_{\rm intra} (\omega)
		&= \frac{1}{\Omega} \frac{ {\rm e}^{-\frac{2S}{\hbar} }}{4\gamma \cos\frac{\theta}{2}} \sum_{n=-\infty}^\infty {}_3\tilde{F}_2\left( \begin{array}{c}\frac{1}{2},1,\frac{3}{2} \\  1+n, 2-n \end{array}; -\gamma^{-2} \right)   \frac{ {\rm e}^{+{\rm i} \pi \left( n_{\rm in}- \frac{1}{2} \right) \left( \frac{\omega}{\Omega} +1 \right)}}{ \frac{\omega+{\rm i}0^+}{\Omega} - (2n-1)} \sum_{\pm} \frac{1}{\cos \left( \frac{\pi}{2} \left( \frac{\omega+{\rm i}0^+}{\Omega} - (2n-1) \pm \frac{\theta}{\pi} \right) \right)}  \label{eqs17} \\
		&= \frac{-{\rm i}}{\Omega} \frac{ {\rm e}^{-\frac{2S}{\hbar} }}{2\gamma \cos\frac{\theta}{2}} \sum_{n=-\infty}^\infty (-1)^n {}_3\tilde{F}_2\left( \begin{array}{c}\frac{1}{2},1,\frac{3}{2} \\  1+n, 2-n \end{array}; -\gamma^{-2} \right) \nonumber\\
			&\quad \times \left[ \frac{\pi}{\cos\frac{\theta}{2}} \delta \left( \frac{\omega}{\Omega} - (2n-1)  \right)  +     \sum_{\pm} \sum_{n'=-\infty}^{+\infty} \frac{ {\rm e}^{+{\rm i} \pi \left( n_{\rm in}-\frac{1}{2} \right) \left( \mp \frac{\theta}{\pi}  -1 \right)}}{\mp \frac{\theta}{\pi}  + (2n'-1) } \delta \left(  \frac{\omega}{\Omega} - 2(n+n'-1) \pm \frac{\theta}{\pi} \right) \right] + ({\rm finite}).   \nonumber
\end{align}
\end{widetext}

Next, we compute $\tilde{J}^\infty_{\rm inter}$.  To perform an analytic calculation, we approximate $\rho$ with a linear function as
\begin{align}
	\rho(t) 
	\sim \frac{\rho(2\pi/\Omega)}{2\pi/\Omega} t 
	= \frac{\theta}{\pi} \hbar \Omega t. \label{eqs18}
\end{align}
Then, the Fourier spectrum $\tilde{J}^\infty_{\rm inter}$ is expressed as
\begin{widetext}
\begin{align}
	\tilde{J}^\infty_{\rm inter} (\omega)
		&\sim  \int^{\infty}_{-\infty} \frac{{\rm d}\omega'}{2\pi} e\tilde{m}(\omega-\omega') \underbrace{  \int^{+\infty}_{-\infty}{\rm d}t\,{\rm e}^{+{\rm i}\omega' t}  (-1)^{-n_{\rm in}+1} {\rm e}^{-\frac{S}{\hbar} } \sum_{n=n_{\rm in}}^{\infty} (-1)^{n} \Theta(t-n \pi/\Omega) \sum_{\pm} {\rm e}^{\pm {\rm i}\frac{\theta}{\pi}(\Omega t-n\pi) } }_{\equiv f_{\rm inter}(\omega')}.   \label{eqs19}
\end{align}
\end{widetext}
We first evaluate $\tilde{m}$.  Noticing that $m$ is an even function in time and has the period $2\pi/\Omega$, one may express $m$ as a Fourier cosine series as
\begin{align}
	m (t) = \frac{c_0}{2} + \sum_{n=1}^\infty c_n \cos (n \Omega t),
\end{align}
where we defined 
\begin{align}
	c_n 
		&\equiv \frac{\Omega}{\pi} \int^{+\pi/\Omega}_{-\pi/\Omega} {\rm d}t\,\cos(n \Omega t) m(t) \nonumber\\
		&= \frac{2}{\pi} \int^{+\pi}_{0} {\rm d}x \frac{\cos(n x)}{\sqrt{ 1 + \gamma^{-2} \sin^2(x)}}.  
\end{align}
We can perform the $x$-integration analytically and eventually finds
\begin{align}
	c_n 	
		= \left\{ \begin{array}{ll}
				0 & (n:\ {\rm odd}) \\
				2 (-1)^{\left\lfloor \frac{n}{2} \right\rfloor}{}_3\tilde{F}_2 \left( \begin{array}{c} \frac{1}{2},\frac{1}{2},1 \\ 1-\frac{n}{2}, 1+\frac{n}{2} \end{array}; -\gamma^{-2} \right) & (n:\ {\rm even})
			\end{array} \right. .
\end{align}
Note that $c_0 = \frac{4}{\pi} {\rm K}(-\gamma^{-2})$.  
Therefore, $\tilde{m}(\omega)$ is evaluated as 
\begin{align}
	\tilde{m}(\omega)
		&= \int^{+\infty}_{-\infty}{\rm d}t\,{\rm e}^{+{\rm i}\omega t} \left[ \frac{c_0}{2} + \sum_{n=1}^{\infty} c_{2n} \cos( 2n \Omega t)   \right]  \nonumber\\
		&= 2\pi \sum_{n=-\infty}^{+\infty}  (-1)^{n}{}_3\tilde{F}_2 \left( \begin{array}{c} \frac{1}{2},\frac{1}{2},1 \\ 1-n, 1+n \end{array} ; -\gamma^{-2} \right) \delta(\omega-2n\Omega)  . \label{eqeq}
\end{align}
Note that $\tilde{m}$ has only even harmonics in contrast to $\tilde{v}$ (\ref{eq106}).  
Let us then evaluate the remaining object $f_{\rm inter}$.  The result reads 
\begin{widetext}
\begin{align}
	f_{\rm inter}(\omega)
		&= \frac{-{\rm i}}{\Omega}{\rm e}^{ +{\rm i} \pi \left( n_{\rm in}  -\frac{1}{2} \right) \frac{ \omega}{\Omega}}  {\rm e}^{-\frac{S}{\hbar} } \frac{\omega/\Omega }{\left(\frac{\omega}{ \Omega} +{\rm i}0^+ \right)^2 - \left(\frac{\theta}{\pi}\right)^2 } \frac{1}{ \cos \left( \pi \left( \frac{\omega}{2\Omega} +{\rm i}0^+ \right)  \right) }  \nonumber \\
		&= \frac{-1}{\Omega} {\rm e}^{ +{\rm i} \pi \left( n_{\rm in}  -\frac{1}{2} \right) \frac{ \omega}{\Omega} }  {\rm e}^{-\frac{S}{\hbar} }\left[ \sum_\pm \delta \left( \frac{\omega}{\Omega} \pm \frac{\theta}{\pi} \right)\frac{\pi}{ \cos \frac{\theta}{2} }    +  2 \frac{\omega/\Omega}{\left( \frac{\omega}{\Omega} \right)^2 - \left( \frac{\theta}{\pi} \right)^2 }\sum_{n=-\infty}^{+\infty} (-1)^n \delta \left( \frac{\omega}{\Omega} - (2n-1)\right)   \right]+ ({\rm finite}).   \label{eqs24} 
\end{align}
\end{widetext}
An important difference from $f_{\rm intra}$ (\ref{eq102}) is that $f_{\rm inter}$ is not peaked at $\omega \sim 0$ but at $|\omega| \sim \frac{\theta}{\pi}$, which enables the interband contribution to contribute to high HHG $|\omega| \sim \frac{\theta}{\pi}$.  

Now, we are ready to compute $\tilde{J}^\infty_{\rm inter}$.  Plugging Eqs.~(\ref{eqeq}) and (\ref{eqs24}) into Eq.~(\ref{eqs19}), we obtain
\begin{widetext}
\begin{align}
	\tilde{J}^\infty_{\rm inter}(\omega)
		&= \frac{-{\rm i}}{\Omega} {\rm e}^{-\frac{S}{\hbar} } \sum_{n=-\infty}^{+\infty}  {}_3\tilde{F}_2 \left( \begin{array}{c} \frac{1}{2},\frac{1}{2},1 \\ 1-n, 1+n \end{array} ; -\gamma^{-2} \right) {\rm e}^{ +{\rm i} \pi \left( n_{\rm in}  -\frac{1}{2} \right) \frac{\omega}{\Omega} } \frac{\frac{\omega}{\Omega}-2n }{\left(\frac{\omega+{\rm i}0^+}{ \Omega} -2n \right)^2 - \left(\frac{\theta}{\pi}\right)^2 } \frac{1}{ \cos \left( \frac{\pi}{2} \left( \frac{\omega+{\rm i}0^+}{\Omega} -2n \right)  \right) }  \nonumber\\
		&= (-1)^{n_{\rm in}} \frac{-{\rm i}}{\Omega} {\rm e}^{-\frac{S}{\hbar} } \sum_{n=-\infty}^{+\infty}  2(-1)^{n}{}_3\tilde{F}_2 \left( \begin{array}{c} \frac{1}{2},\frac{1}{2},1 \\ 1-n, 1+n \end{array} ; -\gamma^{-2} \right)  \label{eqs27} \\
			&\quad \times \left[  \sum_{n'=-\infty}^{+\infty} (-1)^{n'}  \frac{ 2n'-1 }{  (2n'-1)^2 - \left(\frac{\theta}{\pi}\right)^2 } \delta \left(  \frac{\omega}{\Omega} - 2 \left(n+n' \right) +1 \right)  -  \sum_\pm \frac{\pi}{4} \frac{{\rm e}^{ +{\rm i} \pi \left( n_{\rm in}  -\frac{1}{2} \right) (\mp \frac{\theta}{\pi} - 1) }}{ \cos \frac{\theta}{2} } \delta \left( \frac{\omega}{ \Omega} -2n \pm \frac{\theta}{\pi} \right)  \right] + ({\rm finite}).   \nonumber
\end{align} 
\end{widetext}

\subsection{Computation of the Fourier spectrum $\tilde{J}_{\rm obs}(\omega)$ with convoluting a window function}

One can express $\tilde{J}_{\rm obs}(\omega)$ as a convolution integral between $\tilde{J}^\infty_{\rm obs}(\omega)$ and the Fourier transformation for a window function $\tilde{W}(\omega)$ as
\begin{align}
	\tilde{J}_{\rm obs}(\omega) 
		&= \int^{+\infty}_{-\infty} \frac{{\rm d}\omega'}{2\pi} \tilde{W}(\omega-\omega') \tilde{J}^\infty_{\rm obs}(\omega') .
\end{align}
The window function can be anything here and hereafter (for example, $\tilde{W}(\omega) = e^{{\rm i} (t_{\rm in}+T_w/2) \omega } \frac{ \sin \left( \frac{T_w \omega }{2} \right) }{\omega \left( 1 - \left( \frac{T_w\omega}{2\pi} \right)^2 \right)} $ for the Hann window function).  We assume that the dominant contributions to the above integral come from the poles of $\tilde{J}^\infty_{\rm obs}$ and neglect the finite terms in $\tilde{J}^\infty_{\rm obs}$.  Using Eqs.~(\ref{eqs17}) and (\ref{eqs27}), we obtain
\begin{widetext}
\begin{subequations}
\label{es32}
\begin{align}
	\tilde{J}_{\rm intra} (\omega)
		&\sim \frac{-{\rm i}}{2\pi} \frac{ {\rm e}^{-\frac{2S}{\hbar} }}{2\gamma \cos\frac{\theta}{2}} \sum_{n=-\infty}^\infty (-1)^n {}_3\tilde{F}_2\left( \begin{array}{c}\frac{1}{2},1,\frac{3}{2} \\  1+n, 2-n \end{array}; -\gamma^{-2} \right) \\
			&\quad \times \left[ \frac{\pi}{\cos\frac{\theta}{2}} \tilde{W} \left( \omega - (2n-1)\Omega \right)  +  \sum_{\pm} \sum_{n'=-\infty}^{+\infty} \frac{ {\rm e}^{+{\rm i} \pi \left( n_{\rm in}-\frac{1}{2} \right) \left( \mp \frac{\theta}{\pi} -1 \right)}}{\mp \frac{\theta}{\pi}  + (2(n'-n)-1) } \tilde{W} \left(  \omega - (2n'-1)\Omega \pm \frac{\theta}{\pi} \Omega \right) \right] , \nonumber\\
	\tilde{J}_{\rm inter}(\omega)
		&\sim (-1)^{n_{\rm in}} \frac{-{\rm i}}{\pi} {\rm e}^{-\frac{S}{\hbar} } \sum_{n=-\infty}^{+\infty} {}_3\tilde{F}_2 \left( \begin{array}{c} \frac{1}{2},\frac{1}{2},1 \\ 1-n, 1+n \end{array} ; -\gamma^{-2} \right) \\
			&\quad \times \left[  \sum_{n'=-\infty}^{+\infty} (-1)^{n'}  \frac{ 2(n'-n)-1 }{  (2(n'-n)-1)^2 - \left(\frac{\theta}{\pi} \right)^2 } \tilde{W} \left(  \omega - (2n' -1) \Omega \right) -  \sum_\pm (-1)^{n} \frac{\pi}{4} \frac{{\rm e}^{ +{\rm i} \pi \left( n_{\rm in}  -\frac{1}{2} \right) (\mp \frac{\theta}{\pi} - 1) }}{ \cos \frac{\theta}{2} } \tilde{W} \left( \omega -2n\Omega \pm \frac{\theta}{\pi} \Omega  \right)  \right] . \nonumber
\end{align} 
\end{subequations}
\end{widetext}

\subsection{Low frequency limit $\gamma \to 0$}

In the low-frequency limit $\gamma = \frac{(\Delta/2)\Omega}{eE_0} \to 0$, the Fourier spectrum (\ref{es32}) asymptotes
\begin{widetext}
\begin{subequations}
\label{eqs33}
\begin{align}
	\tilde{J}_{\rm intra} (\omega)
		&\to \frac{{\rm i}}{\pi} {\rm e}^{-\frac{2S}{\hbar} } \frac{1}{\cos^2\frac{\theta}{2}}\sum_{n=-\infty}^\infty \left[ \frac{\tilde{W} \left( \omega - (2n+1)\Omega \right) }{2n+1}  -   \sum_{\pm}  {\rm e}^{+{\rm i} \pi \left( n_{\rm in}-\frac{1}{2} \right) \left( \mp \frac{\theta}{\pi}  -1 \right)} \frac{ \sin\frac{\theta}{2}}{4\left(n-\frac{\theta}{2\pi}\right)}  \tilde{W} \left(  \omega -\Omega \mp 2\left(n-\frac{\theta}{2\pi}\right) \Omega \right) \right]  ,  \\
	\tilde{J}_{\rm inter}(\omega)
		&\to (-1)^{n_{\rm in}} \frac{-{\rm i}}{\pi} {\rm e}^{-\frac{S}{\hbar} } \frac{\gamma}{4\cos\frac{\theta}{2}} \sum_{n=-\infty}^{+\infty} \sum_\pm  \left[ \left( {\rm ln}\,\gamma^2 + 2H_{\pm n-1/2} \right) {\rm e}^{ +{\rm i} \pi \left( n_{\rm in}  -\frac{1}{2} \right) \left(\mp \frac{\theta}{\pi} - 1 \right) } \tilde{W} \left( \omega \mp 2\left(n-\frac{\theta}{2\pi}\right)\Omega \right)  \right] \nonumber\\
			&\quad \left. + \left( {\rm ln}\,(4\gamma^2) - \left( (-1)^n \cos\frac{\theta}{2} - 1  \right) H_{\frac{n}{2}\pm \frac{\theta}{4\pi}-1} + \left(  (-1)^n \cos\frac{\theta}{2} + 1  \right) H_{\frac{n}{2} \pm \frac{\theta}{4\pi}-\frac{1}{2}} \right)  \tilde{W} \left(  \omega - (2n -1) \Omega \right) \right]    , 
\end{align} 
\end{subequations}
\end{widetext}
which is Eq.~(\ref{eq_10}) in the main text, and $S$ and $\theta$ are expanded in terms of $\gamma$ as
\begin{align}
	S = \frac{\pi}{2} \nu \gamma^2 \left[ 1 + {\mathcal O}(\gamma^2) \right],\ 
	\theta = 4\nu \left[ 1 + {\mathcal O}(\gamma^2) \right] . 
\end{align}
Note that $|\beta|^2 \sim {\rm e}^{-2S/\hbar}$ characterizes the typical magnitude of the pair production number, and $ {\rm e}^{-2S/\hbar} \to \exp \left[ -\pi \frac{(\Delta/2)^2}{\hbar eE_0}  \left[ 1+{\mathcal O}(\gamma^2) \right] \right]$ ($\gamma \to 0$) reproduces the non-perturbative Schwinger formula (or the Landau-Zener formula) driven by quantum tunneling.

\bibliography{bib}

\begin{thebibliography}{91}
\bibitem [{foo()}]{foot1}%
  \BibitemOpen
  \bibinfo {note} {When an applied electric field is spatially uniform and
  linearly polarized, spin degrees of freedom are degenerated no matter what
  the spacetime dimension is and hence the Dirac Hamiltonian reduces in general
  to a two-level system, describing the valence- and conduction-band-electrons
  degrees of freedom.}\BibitemShut {Stop}%
\end{thebibliography}

\end{document}